\documentclass[aps,prl,reprint,superscriptaddress]{revtex4-2}
\usepackage[inline]{enumitem}
\usepackage{amsmath}
\usepackage{graphicx}
\usepackage{dcolumn}
\usepackage{bm}
\usepackage{color}
\usepackage{placeins}
\usepackage{hyperref}
\usepackage{orcidlink}
\hypersetup{
 colorlinks,
 linkcolor={red},
 citecolor={blue},
 urlcolor={blue}
}
\usepackage{xcolor}
\usepackage{xfrac}
\usepackage[ulem=normalem]{changes}
\usepackage{mathpazo}
\usepackage{framed}
\usepackage{multibib}
\usepackage{float}

\setlist[itemize]{noitemsep, nosep, itemsep=3pt, topsep=3pt, itemindent = -1em}
\setlist[enumerate]{noitemsep, nosep, itemsep=3pt, topsep=3pt}

\makeatletter
\renewcommand{\p@subsection}{}
\renewcommand{\p@subsubsection}{}
\makeatother

\begin{document}

\title{Automated Physics-Informed Neural-Networks-Based Calibration of Highly Segmented Silicon Telescopes}

\author{M.~Rejmund\,\orcidlink{0009-0009-8626-8756}}
\affiliation{GANIL, CEA/DRF - CNRS/IN2P3, Bd Henri Becquerel, BP 55027, F-14076 Caen Cedex 5, France}
\email[M. Rejmund: ]{mrejmund@ganil.fr}
\author{A.~Lemasson\,\orcidlink{0000-0002-9434-8520}} 
\affiliation{GANIL, CEA/DRF - CNRS/IN2P3, Bd Henri Becquerel, BP 55027, F-14076 Caen Cedex 5, France}
\author{P.~Morfouace\orcidlink{0000-0002-2131-2199}} 
\affiliation{CEA, DAM, DIF, F-91297, Arpajon, France}
\affiliation{Universit\'e Paris-Saclay, CEA, LMCE, F-91680, Bruy\`eres-le-Ch\^atels, France}

\author{D.~Ramos\,\orcidlink{0000-0003-1822-6537}} 
\affiliation{GANIL, CEA/DRF - CNRS/IN2P3, Bd Henri Becquerel, BP 55027, F-14076 Caen Cedex 5, France}

\author{J.~Taieb} 
\affiliation{CEA, DAM, DIF, F-91297, Arpajon, France}
\affiliation{Universit\'e Paris-Saclay, CEA, LMCE, F-91680, Bruy\`eres-le-Ch\^atels, France}

\author{J.~D.~Frankland\,\orcidlink{0000-0002-4907-5041}} 
\affiliation{GANIL, CEA/DRF - CNRS/IN2P3, Bd Henri Becquerel, BP 55027, F-14076 Caen Cedex 5, France}

\date{\today}

\begin{abstract}
Transfer and multi-nucleon transfer reactions are essential tools for probing nuclear structure and reaction dynamics, 
requiring precise determination of the particle identification, energy, and emission angles of reaction products. The increasing granularity 
of modern silicon telescope arrays significantly enhances experimental capabilities but introduces major challenges for 
detector calibration, as conventional channel-by-channel approaches become inefficient, difficult to scale, and potentially 
inconsistent.

In this work, we present a fully automated, physics-informed calibration framework based on neural networks, specifically 
designed for highly segmented silicon detector arrays. The method formulates calibration as a global optimization problem, 
in which detector gains and geometrical corrections are determined simultaneously by minimizing the width of the reconstructed 
excitation energy under two-body kinematics constraints. The approach relies exclusively on experimental data and 
well-established physical principles, without requiring explicit modeling of detector response.

A distinctive feature of the method is the use of multiple neural network sub-models sharing a common loss function
with embedded physics constraints, enabling coherent and self-consistent calibration across all detector 
channels. This strategy ensures scalability, robustness, and reproducibility, making it particularly suitable for next-generation 
detector systems with increasing complexity.

The performance of the method is demonstrated using experimental data from the Particle-Identification Silicon-Telescope 
Array (PISTA) in high-resolution fission studies in inverse kinematics. The results show excellent agreement with predicted 
kinematics, high-quality particle identification, and a significant improvement in calibration efficiency. The proposed framework 
provides a general and adaptable solution for the calibration of complex detector systems in modern nuclear physics experiments.

\end{abstract}

\maketitle

\section{Introduction}
\label{sec:intro}

Transfer reactions and, more generally, multi-nucleon transfer reactions constitute a powerful tool 
for investigating nuclear structure and reaction mechanisms. They provide direct access to single-particle 
and collective properties and their evolution across the nuclear chart~\cite{wimmer2018}.
The experimental study of such processes requires precise determination of the particle identification, energy, and emission 
angle of reaction products, placing stringent constraints on the performance of detection systems. 
In this context, silicon detector telescopes have become a cornerstone of modern nuclear physics experiments 
due to their excellent energy resolution, particle identification capabilities, and fast response.

Over the past decades, continuous developments in detector technology have led to increasingly sophisticated 
silicon arrays, characterized in particular by a high degree of segmentation. Single or double-sided strip detectors and 
pixelated geometries now allow for fine angular resolution and large solid-angle coverage, enabling detailed 
reconstruction of complex reaction kinematics. 
Modern detection systems may comprise hundreds or even thousands of individual channels, each 
providing localized information on the measured physical observables.

While such segmentation significantly enhances the experimental capabilities, it also introduces new challenges 
in data analysis, and in particular in detector calibration. In highly segmented systems, each individual channel 
samples only a limited fraction of the overall physical phase space and often results in reduced statistics 
per segment. Traditional calibration procedures, typically performed on a channel-by-channel basis through iterative and 
often manual adjustments, become increasingly difficult to implement efficiently in this context. These 
approaches are not only time-consuming but may also introduce subjective biases and lack global consistency 
across the detector array.

The rapid growth in the number of channels and the complexity of the measured observables therefore call 
for the development of new, more robust calibration strategies. In particular, there is a need for methods 
capable of handling high-dimensional data, exploiting correlations between detector segments, and providing 
reproducible and scalable solutions. In this respect, machine learning techniques, and neural networks in 
particular, offer a promising alternative to conventional approaches. By construction, neural networks are 
well suited to model complex, non-linear relationships in multi-parametric datasets, and can be trained 
to extract optimal calibration mappings from experimental or simulated data.

In this work, we propose a fully data-driven, globally constrained calibration method using physics-informed 
loss functions tailored to highly segmented silicon telescope arrays employed in transfer reaction studies.

In neural-network training, a loss function quantifies the discrepancy between the desired physical behavior and 
the result predicted by the network. During optimization, the network parameters are adjusted to minimize this 
quantity. In the present work, the physics-informed loss function is constructed analytically from the deviation 
of the reconstructed reaction kinematics from the known physical constraint.

The primary objective of this work is  to provide a time-efficient and fully reproducible procedure, relying on a 
mathematically well-defined optimization process. Furthermore, its inherent scalability makes it particularly well-suited for 
next-generation detector systems characterized by ever-increasing granularity.
The distinguishing features of our method are: 
\textbf{(i)} the exclusive utilization of available data and the 
well-established two-body reaction kinematics, and \textbf{(ii)} the utilization of a neural network model 
composed of three independent sub-models that share a common loss function in which the physics 
constraints were embedded.

By leveraging the ability of neural networks to capture global correlations across detector channels, this method 
offers a consistent and flexible solution to the challenges posed by modern multi-parametric nuclear physics 
experiments.

Our approach is demonstrated through an application to experimental data obtained by the highly 
segmented Particle-Identification Silicon-Telescope Array (PISTA)~\cite{BegueGuillou2026} in the 
high-resolution studies of fission processes induced in inverse kinematics.

Beyond the specific application to PISTA, the work demonstrates a general strategy for detector calibration in 
which experimental data are combined with known physical constraints to determine multiple detector and 
geometrical parameters simultaneously. The neural-network framework provides a flexible and scalable means 
of performing this optimization, while the resulting parameters can be directly used as conventional calibration 
constants in subsequent analyses. This approach is particularly relevant for highly segmented detector systems, 
for which channel-by-channel calibration and independent determination of geometrical parameters become increasingly 
demanding as the number of channels grows.

\section{PISTA array}
\label{sec:PISTA}

\begin{figure*}[ht]\centering
\includegraphics[width=0.9\textwidth]{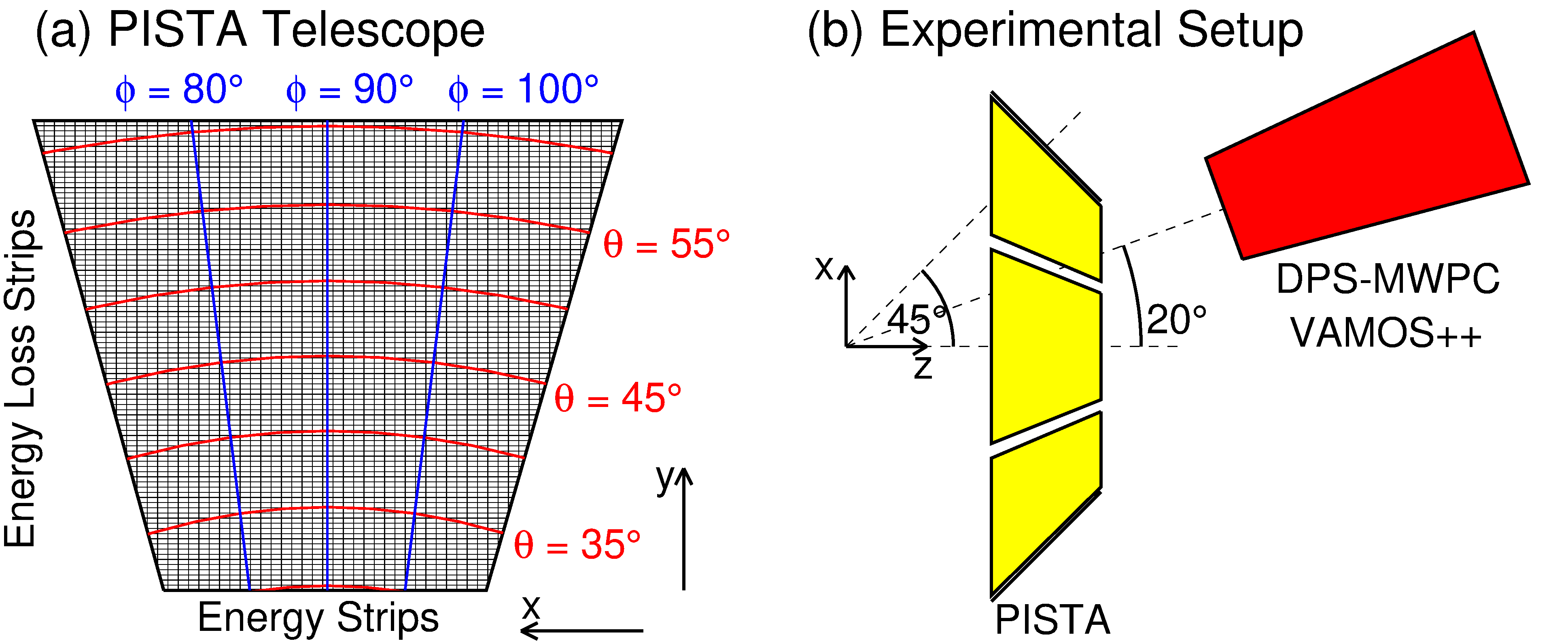}
\caption{PISTA array and experimental setup:
(a) The schematic front view of the segmentation of the trapezoidal telescope of PISTA located at 
the azimuthal angle $\phi = 90^\circ$.
This diagram illustrates 91 horizontal strips of the $\Delta E$ stage and 57 vertical strips of the $E_{res}$ stage.
In the reference frame of the telescope, the $z$ axis aligns with the telescope's normal.
The origin is positioned at $105.5$~mm, and the $z$ axis intersects the detector’s plane at 
the centers of the middle horizontal and vertical strips.
The red lines represent the coordinates for the constant polar laboratory angle $\theta$ in 
increments of 5 degrees, while the blue lines indicate those for the constant azimuthal angle 
$\phi$ in increments of $10$ degrees.
(b) The schematic top view of the experimental setup. The $^{12}\rm{C}$ target is positioned in the $xy$ plane 
at the origin of the coordinate system, while the $^{238}\rm{U}$ beam axis aligns with the $z$ axis.
The PISTA telescopes encircle the target at a distance of $105.5$~mm and a polar angle 
of $\theta = 45^\circ$. The corresponding telescope’s normal is indicated by dashed line. Eight telescopes 
are placed at the azimuthal angle $\phi$ every $45^\circ$.
The dual position-sensitive Multi-Wire Proportional Chamber (DPS-MWPC)
placed at the entrance of the VAMOS++ spectrometer, rotated in the $xz$ plane by $20^\circ$ relative to the 
beam axis is also shown; its normal is indicated by dashed line.
Inverse kinematics is exemplified by the $^{238}$U+$^{12}$C reaction. The target-like $^{12}$C is detected in 
the PISTA array, providing its energy and emission angle, while the fission of the projectile-like $^{238}$U produces 
two fission fragments, of which one ($\mathrm{FF}_1$) is detected and identified in the VAMOS++ spectrometer. 
The measured properties of the detected target-like product are used to reconstruct the reaction kinematics.
\label{fig:PISTA}}
\end{figure*}

The PISTA array~\cite{BegueGuillou2026} was designed to identify light reaction products and determine 
their emission angles in inverse-kinematics reactions. In the present experiment, a $^{238}$U beam 
impinges on a $^{12}$C target, producing reaction products that are detected in PISTA, while one of 
the fission fragments is identified and tracked in the VAMOS++ spectrometer~\cite{Pullanhiotan2008, Rejmund2011}. 
The measured energy and emission angle of the target-like particle, together with the fission-fragment information, allow 
the excitation energy of the fissioning system to be reconstructed. Accurate calibration of the PISTA 
energy and position measurements is therefore essential for achieving the required excitation-energy resolution.

PISTA comprises eight trapezoidal telescopes, each equipped with two stages, $\Delta E$ and $E_{res}$, 
of single-sided stripped silicon detectors. The detectors have thicknesses of typically $100~\mu$m and 
$1$~mm, respectively.
The $\Delta E$ detectors have $0.5~\mu$m-thick aluminum entrance and exit windows, while the $E_{\rm res}$ detectors 
have a $0.5~\mu$m-thick aluminum entrance window.

Figure~\ref{fig:PISTA}(a) illustrates the schematic representation 
of the trapezoidal telescope within the coordinate frame of the telescope.
The front side of the first stage ($\Delta E$) is divided into $91$ 
horizontal strips, each measuring $535~\mu$m in width, separated by $100~\mu$m of passive inter-strips. 
The front side of the second stage ($E_{res}$) is divided into $57$ vertical strips, each measuring $1.17$~mm 
in width, separated by $100~\mu$m of passive inter-strips. As illustrated in Figure~\ref{fig:PISTA}(a), the trapezoidal 
shape of the telescope causes the lengths 
of the $\Delta E$ and $E_{res}$ strips to vary. The distance between the first stage of the telescope to the target 
is $105.5$~mm while the distance between the first and the second stage is $4.4$~mm. The PISTA telescopes
are positioned at the polar angle of $\theta = 45^\circ$ relative to the beam axis. The coordinates corresponding 
to the constant polar laboratory angle $\theta$ are depicted in red, while those for the constant azimuthal 
laboratory angle $\phi$ are shown in blue. From the figure, it can be inferred that a single $\Delta E$ and 
$E_{res}$ strip covers a range of laboratory angles $\theta$ and $\phi$, respectively.

Figure~\ref{fig:PISTA}(b) illustrates the top view of the experimental configuration. The $z$ axis is aligned with 
the beam axis, forming a horizontal plane with the $x$ axis, while the $y$ axis is aligned vertically. 
The target, situated at the origin of the coordinate system, is encircled by the eight telescopes 
of the PISTA array. These telescopes are positioned at a polar angle of $\theta = 45^\circ$ and an azimuthal 
angle $\phi$ that increments by $45^\circ$ commencing from $\phi = 0^\circ$.
The complete PISTA array contains $8\times(91+57)=1184$ strip channels. 

The VAMOS++ spectrometer is situated downstream of the beam axis 
and rotated by $20^\circ$ about the $y$ axis. Its entrance detector, the dual position-sensitive Multi-Wire 
Proportional Chamber (DPS-MWPC)~\cite{Vandebrouck2016}, is positioned at a distance of $167$~mm from 
the target and is depicted in the figure.

Further,  the schematic of the experimental principle is illustrated in Figure~\ref{fig:PISTA}(b). The reactions in inverse kinematics 
are induced by the interaction of the $^{238}$U beam with the $^{12}$C target. The target-like reaction products, 
such as the inelastically scattered $^{12}$C in this example, are detected by the PISTA array. The beam-like products, 
such as $^{238}$U, can undergo fission. In this case, only one fission fragment, FF$_1$, is detected in the VAMOS++ 
spectrometer. Conversely, in the case of elastic scattering, the outgoing $^{238}$U does not undergo fission and remains 
undetected.

\section{Experimental methods}
\label{sec:EXP}

The experimental data utilized in this study were obtained during the E850 GANIL experiment~\cite{DataE850}, 
which was designed to detect, identify, and track light nuclei in coincidence with fission fragments produced 
in multi-nucleon transfer reactions in inverse kinematics. The reactions were induced by a $^{238}$U beam
with an energy of $5.95$~MeV/u on a $100~\mu$g/cm$^2$ thick $^{12}$C target. 
The fission fragments were detected and isotopically identified in VAMOS++ spectrometer.
Light nuclei were detected
in PISTA array. 

\subsection{PISTA methods}
\label{sec:PISTAMeth}

For the measurement of the energy loss $\Delta E$, the individual front strips from the first 
stage of the PISTA telescopes were utilized. Conversely, for the measurement of the residual energy $E_{res}$, 
the unstripped rear sides of the second stage were employed. This selection was guided by the optimal energy 
resolutions corresponding to each measurement. As previously mentioned, the trapezoidal shape of the telescope 
causes the length of the strips to vary. Consequently, due to capacitive coupling with the strips, the gain of the 
rear side stage may differ from one strip to another. Therefore, its energy calibration was assumed to vary as a 
function of the position of 
the corresponding $E_{res}$ strip. To determine the position of the interaction of the nucleus with the 
detector, the individual front strips from the first stage and the second stage were utilized. The centers 
of the corresponding strips were utilized as the defining point of interaction.

One of the key assets of the PISTA array is its high granularity, which is designed to guarantee a high-resolution 
measurement of the polar angle $\theta$. This enables the high-resolution measurement of the excitation energy 
of the fissioning system ($E^*$). Towards this goal, the interaction point of the beam with the target has to be 
determined, as the standard deviation of the typical beam profile is comparable to the $\Delta E$ strip width.
The event-by-event measurement of the interaction position on the target is provided by the DPS-MWPC placed 
at the entrance of VAMOS++ spectrometer and illustrated in Fig.~\ref{fig:PISTA}(b). However, this measurement 
was only available for the events where the fission fragment was detected in VAMOS++.
For the elastic scattering of $^{238}$U on $^{12}$C, which was used for the calibration of the telescopes, 
the interaction position on the target was not available. The mean beam position $(\overline{x}, \overline{y})_{b}$, 
corresponding to the average interaction position determined over the last $5000$ fission events, was used.

In this experiment, to enhance the production of the desired nuclei, a high beam current was used, 
leading to an overwhelming rate of elastically scattered $^{12}$C nuclei impinging on PISTA at the 
larger laboratory angles. Hence the outermost $\Delta E$ strips were masked, leaving approximately 
$45$ inner $\Delta E$ strips active.

To illustrate the application of our calibration method, we exclusively utilized a single initial data run of the 
experimental data. This data were utilized to minimize the width of the excitation energy of the system ($E^*$) for 
the elastic scattering of $^{238}$U on $^{12}$C, assuming that the scattering occurred in the middle of the target. 
The mean beam position during the run was $(\overline{x}, \overline{y})_{b} = (0.06, 0.05)_{b}$~mm. 
The corresponding width of the mean beam position due to its movement on the target was 
$(\sigma(\overline{x}), \sigma(\overline{y}))_{b} = (0.03, 0.04)_{b}$~mm. 
In conjunction with these data, the pre-experiment 
three-alpha calibration data available for the $\Delta E$ stage were utilized.
The primary focus was on three strategically positioned telescopes at the azimuthal angles 
$\phi = 90^\circ$, $135^\circ$, and $180^\circ$.

\begin{figure*}[th]\centering
\includegraphics[width=0.45\textwidth]{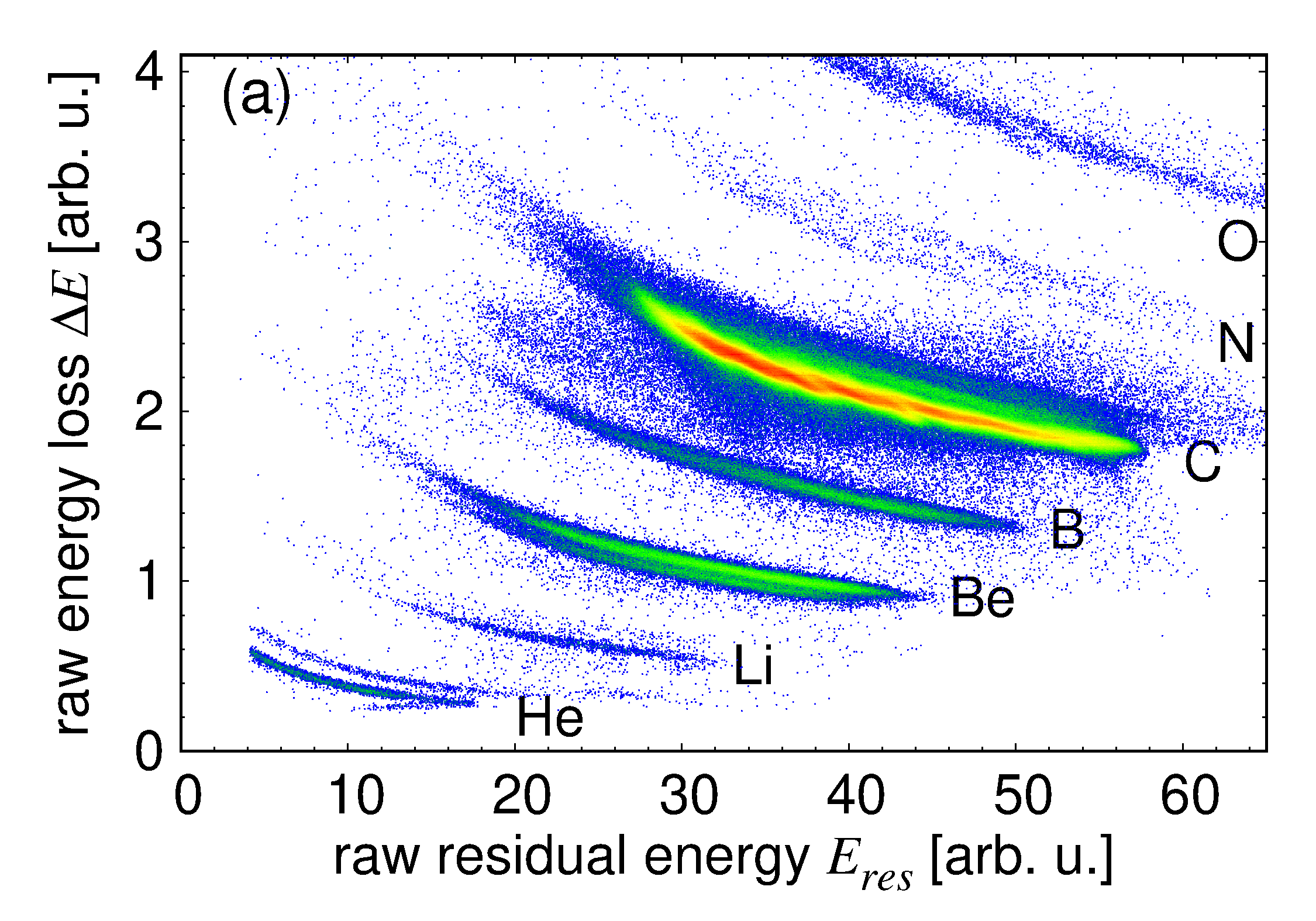}
\includegraphics[width=0.45\textwidth]{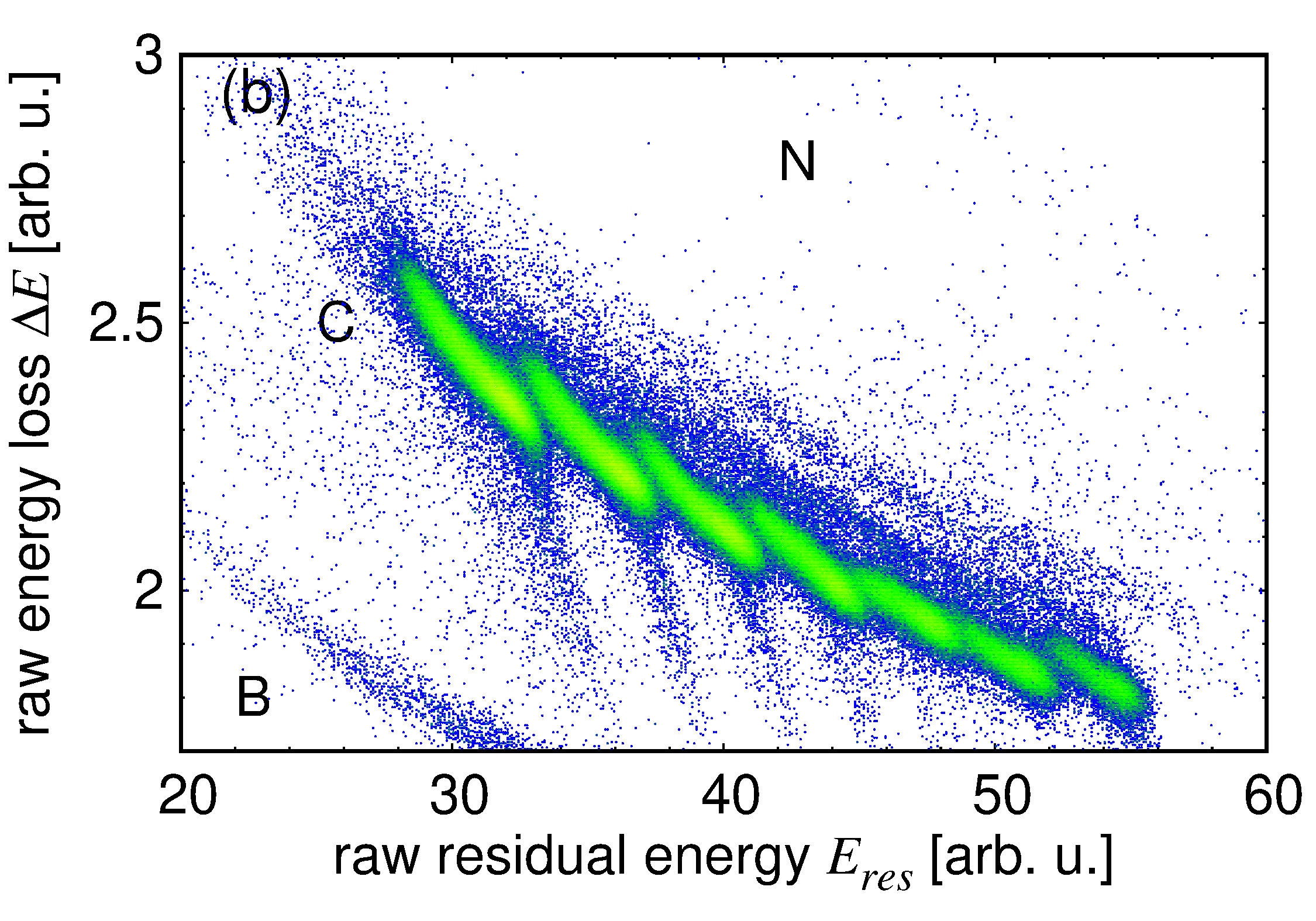}
\caption{Initial identification charts:
Two-dimensional energy correlation plots between the raw energy loss ($\Delta E$) and the raw residual 
energy ($E_{res}$), resulting in correlated bands for each atomic number and atomic mass number. 
(a) Complete $\Delta E$ and $E_{res}$ correlation, including the elements: He, Li, Be, B, C, N, and O, registered 
by the telescope located at $\phi = 135^\circ$.
(b) Zoom on the isotopes of C, where only the data from every sixth $\Delta E$ strip are shown to 
highlight the $\Delta E$ energy range registered by the individual strips. 
\label{fig:dEERAW}}
\end{figure*}

\subsubsection{Raw data}
\label{sec:RawData}

Figure~\ref{fig:dEERAW}(a) illustrates the initial uncalibrated identification chart obtained from the telescope 
situated at $\phi = 135^\circ$. It depicts the two-dimensional energy correlation plot between the raw energy 
loss ($\Delta E$) and the raw residual energy ($E_{res}$), where the correlated bands correspond to distinct 
atomic numbers and atomic mass numbers. The raw energy loss was derived from the front segmented side 
of the $\Delta E$ stage, while the raw residual energy was obtained from the unsegmented rear side of the 
$E_{res}$ stage. Figure~\ref{fig:dEERAW}(b) presents a zoomed-in view of the isotopes of carbon.
To emphasize the energy loss data range collected by individual strips, only the data for every sixth $\Delta E$ 
strip are presented. The most prominent correlation observed for $^{12}$C corresponds predominantly to the 
process of elastic scattering of $^{238}$U on $^{12}$C. These elastic scattering data were utilized in this work 
in conjunction with the pre-experiment three-alpha calibration data available for the $\Delta E$ stage only for 
the calibration of the PISTA telescopes.

\subsubsection{PISTA's resolution and sensitivity}
\label{sec:PISTARES}

In Ref.~\cite{BegueGuillou2026}, the individual contributions to the excitation energy resolution of the PISTA array 
for the elastic scattering of $^{238}$U on $^{12}$C were determined using Monte Carlo simulations. The resulting widths, 
expressed in terms of the standard deviation, are as follows:
\begin{itemize}
\item PISTA’s granularity: $0.124$~MeV,
\item intrinsic energy resolution: $0.126$~MeV,
\item target thickness: $0.208$~MeV, and
\item beam spot profile width: $0.416~(46)$~MeV.
\end{itemize}
The beam profile was determined as:
$\sigma(b)_x = 350 \pm 50~\mu$m and $\sigma(b)_y = 550\pm50~\mu$m
using the DPS-MWPC. The position resolution of the DPS-MWPC on the target was approximately $\sigma_{x,y} \sim 400~\mu$m.
From the individual contributions to the excitation energy resolution of PISTA, it is evident that the dominant 
is related to the beam profile. The total mean simulated excitation energy resolution yielded $\overline{\sigma_s}_{E^*} = 0.500(39)$~MeV, 
which compares with the experimentally obtained $\overline{\sigma}_{E^*} = 0.518(1)$~MeV.
Since the beam profile has different widths in vertical and horizontal directions, the telescopes placed at different
azimuthal angles will exhibit a different excitation energy resolution. From the same simulation as in Ref.~\cite{BegueGuillou2026}
one obtains for the different telescopes: 
$\sigma_s(\phi = 90^\circ)_{E^*} = 0.563(39)$~MeV, $\sigma_s(\phi = 135^\circ)_{E^*} = 0.500(39)$~MeV, and 
$\sigma_s(\phi = 180^\circ)_{E^*} = 0.429(39)$~MeV.
By comparing the beam vertical and horizontal widths, and the excitation energy width in the telescopes placed at 
$\phi = 90^\circ$ and 
$\phi = 180^\circ$ it can be inferred that an increase in the standard deviation of the beam profile by 
$100~\mu$m results in an approximate increase of $0.067$~MeV in the standard deviation of the excitation energy resolution. 
Consequently, the PISTA array demonstrates a higher sensitivity to variations in the beam profile width compared to the DPS-MWPC.

For the reactions where one fission fragment is detected in VAMOS, the position of the interaction of the beam with the target 
is determined on an event-by-event basis, and the excitation energy resolution is limited by the position resolution.
The position resolution is mainly governed by the position resolution of the DPS-MWPC.

This demonstrates that the achievable excitation energy resolution is fundamentally limited by the beam profile and the beam profile 
reconstruction rather than detector performance.

\section{Application of deep neural networks}
\label{sec:PISTANN}

The objective of this work was to achieve the calibration 
of the highly segmented telescopes within the PISTA array. This calibration enables the high-resolution 
measurement of the excitation energy of the populated system and precise particle identification in terms 
of atomic number and atomic mass number.

Traditional methods such as channel-by-channel fits may not demonstrate global consistency. In contrast, 
the polynomial global fit exhibits poor scalability. The neural network approach, on the other hand, effectively 
handles non-linearity, shared constraints, and is easily scalable. 

For the neural-network calibration, the experimental reaction data were selected using the scattered 
$^{12}$C component of the $^{238}$U+$^{12}$C reaction. These events provide the two-body kinematic constraint 
used in the physics-informed loss function. The three-$\alpha$ source data were used to anchor the absolute energy scale
of the $\Delta E$ strips. Other reaction products visible in the measured $\Delta E$--$E_{\rm res}$ 
spectra were not used as constraints in the calibration.
The $^{12}$C events were selected from the corresponding locus in the measured $\Delta E$--$E_{\rm res}$ 
distributions; events associated with other reaction products were therefore excluded from the physics-informed 
calibration sample.

To accomplish the objective of calibration, a regression-type neural network model~\cite{kinsley2020, Subasi2020} 
was employed. We refer to the neural-network calibration model developed in this work as PISTANN (PISTA Neural Network).
This model provided:
\begin{itemize} 
\item individual gains ($a_k$) for each $\Delta E$ strip ($k_{\Delta E}$), 
\item the gain ($b_l$) of the $E_{res}$ detector as a function of the corresponding strip number ($l_{E_{res}}$), 
\item corrections to the telescope’s spatial placement relative to the theoretical one.
\end{itemize}
Precise knowledge of the telescope’s position is crucial to obtain the correct correlation between the 
measured total energy ($E_{tot}$) of the nucleus and its emission polar angle ($\theta$), thereby achieving the 
highest possible resolution of the measured excitation energy of the populated system. 
The most pertinent and constraining corrections to the telescope’s spatial placement are those related to the 
polar angle, expressed as:
\textbf{(i)} the vertical shift ($\delta y$) in the coordinate frame of the detector relative to its normal 
(Figure~\ref{fig:PISTA}(a)),
\textbf{(ii)} the polar angle shift ($\delta\theta$) in the laboratory frame (Figure~\ref{fig:PISTA}(b)).

Silicon detectors exhibit dead layers in both the junction and ohmic regions. 
In PISTA those correspond to the entrance and exit $0.5~\mu$m-thick aluminum windows. 
When the calibration obtained 
with carbon nuclei is applied to other nuclei, differences in the energy loss within these dead layers may 
become non-negligible compared with the excitation-energy resolution. In such cases, energy-loss models 
can be used to account for these effects.
One approach consists of explicitly correcting the measured total energy during training by adding the 
calculated energy loss of carbon in the dead layers as a function of the incidence angle. The excitation 
energy of other nuclei can then be reconstructed by applying the corresponding dead-layer corrections 
calculated for those nuclei.
Alternatively, the energy loss of carbon in the dead layers can be absorbed into the calibration coefficients 
determined by the neural network. In this case, the excitation energy of other nuclei is reconstructed by 
applying only the calculated difference between their dead-layer energy loss and that of carbon.
Since both approaches are mathematically equivalent up to the treatment of the dead-layer correction, 
neither affects the achieved excitation-energy resolution.
In the present work, the latter approach 
was adopted because it allows straightforward testing and comparison of different energy-loss models.

To achieve the objective of calibration, the neural network model minimizes the width of the reconstructed excitation energy, 
obtained employing the two-body kinematics, and derived for the elastic scattering process of 
$^{238}$U on $^{12}$C. The excitation energy $E^*=0$ was assumed for the elastic scattering. 
The minimization is evaluated in terms of the root mean square deviation (RMSD). Furthermore, 
the model concurrently minimizes the widths obtained from the pre-experiment three-alpha calibration data 
available for the $\Delta E$ strips. The role of the three-alpha calibration in the model is to provide an anchor 
for the training process.
Its contribution to the overall width is small since there is about a factor $10$ in widths 
between the alpha and the excitation energy data.
Additionally, the three-alpha calibration data are used to subtract the offsets, in the units of channels, 
for the $\Delta E$ data.

The calibration problem is intrinsically under-constrained when considering gains and geometrical corrections independently. 
In particular, partial degeneracies may arise between detector gains and angular corrections. 
For instance, a systematic increase in gain can be partially compensated by a shift in the reconstructed angle, leading to similar 
excitation energy distributions. These degeneracies are effectively 
lifted by the simultaneous use of alpha calibration data, which anchors the absolute energy scale, and the two-body kinematics 
constraint, which enforces a global correlation between energy and emission angle.

The neural network is used to optimize the calibration and geometrical parameters of the detector. 
Once the training has converged, the optimized parameters are extracted and used as calibration constants 
in the subsequent standard data analysis. Thus, the neural network does not need to be applied event-by-event after 
calibration. In the present case, the physics constraint is provided by the known kinematics of elastic $^{238}$U+$^{12}$C 
scattering, for which the target excitation energy is $E^*_{\rm target}=0$.

The loss function implements the analytical two-body kinematics, which relates the excitation energy to the measured 
total energy and scattering angle, $E^*=E^*(E_{\rm tot},\theta_L)$. Here, $E_{\rm tot}$ is calculated from the raw detector 
signals and the learned energy-gain parameters, $E_{\rm tot}=E_{\rm tot}(\Delta E_k^{\rm raw},E_{{\rm res},l}^{\rm raw},a_k,b_l)$, 
while the laboratory scattering angle is obtained from the detector geometry and the learned geometrical parameters, 
$\theta_L=\theta_L(\theta_{\rm geom},\delta y,\delta\theta)$. The calibration parameters $a_k$, $b_l$, $\delta y$, and 
$\delta\theta$ are optimized by minimizing the deviation of the reconstructed excitation energy from the known target 
value, $(E^*-E^*_{\rm target})^2$, through the corresponding loss function. The gradients of this loss with respect to the 
learned parameters are used for the optimization.

More generally, the same approach can be applied by replacing the present kinematic constraint with the appropriate 
known physical relation for the reaction and detector configuration under study, such as energy or momentum conservation, 
a known reaction $Q$ value, or an invariant-mass constraint.

In summary, the learned parameters are gains $a_k$, $b_l$, and placement corrections $\delta y$ and $\delta\theta$. 
The imposed parameters are two-body kinematics and intrinsic geometry.

Consequently, the neural network does not learn the physical laws themselves, which are explicitly imposed through the loss function, 
but rather infers the detector response parameters that best satisfy these constraints. 

\begin{figure*}[ht]\centering
\includegraphics[width=0.9\textwidth]{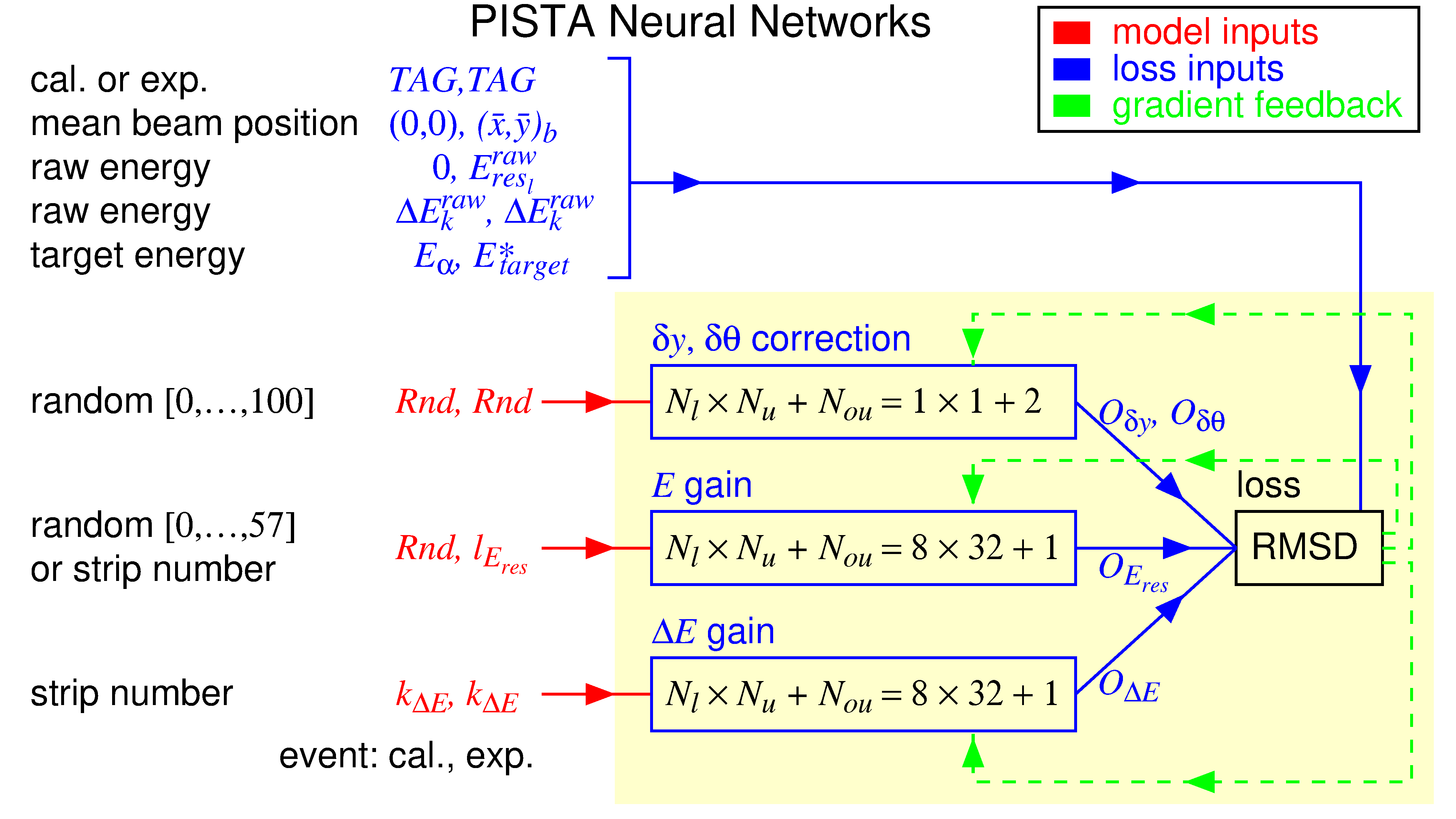}
\caption{PISTA neural networks: Neural networks model consisting of three independent sub-models
trained to provide $\Delta E$ and $E_{res}$ gains along with telescope placement corrections in terms of 
the vertical shift $\delta y$ and the polar angle shift $\delta\theta$.
The architecture of each of the sub-models 
is given in the form  $N_l \times N_u + N_{ou}$ comprising $N_l$ layers and $N_u$ units (neurons) per layer, 
followed by $N_{ou}$ output units. The inputs of the model are indicated in red, the inputs to the loss
function are indicated in blue and the gradient's feedback obtained for a loss are indicated in green.
\label{fig:MODEL}}
\end{figure*}

\subsection{Calibration model}
\label{sec:CalibModel}

\subsubsection{Model formalization}
\label{sec:Form}

The inverse problem addressed by the PISTA neural networks model can be formally expressed as follows:
\begin{itemize}
\item Observables: 
$(\Delta E_k^{raw}$, $E_{res_l}^{raw}$, $\theta_{geom}$),
\item Calibration parameters: 
$a_k$, $b_l$, $\delta y$, $\delta\theta$, 
\item Mapping: 
($\Delta E_k^{raw}$, $E_{res_l}^{raw}$, $\theta_{geom}$) $\rightarrow$ ($E_{tot}, \theta_L)$ $\rightarrow$ $E^*$,
\item Objective function: 
$\mathcal{L} = {\rm RMSD}(E^* - E^*_{target})$.
\end{itemize}

\subsubsection{Sub-models}
\label{sec:Sub}
Figure~\ref{fig:MODEL} illustrates schematically the PISTA neural networks model used. 
The model consists of three independent sub-models: 
\begin{itemize}
\item $\Delta E$ gain ($a_k$) as a function of the strip number ($k_{\Delta E}$), 
\item $E_{res}$ gain ($b_l$) as a function of the strip number ($l_{E_{res}}$),
\item placement corrections in terms of the vertical shift ($\delta y$), and the polar angle shift 
($\delta\theta$) as constants for each telescope.
\end{itemize}
The independence of the sub-models ensures the independence of the obtained gains for $\Delta E$ and 
$E_{res}$, as well as the constancy of the $\delta y$ and $\delta\theta$ corrections.
The general form of neural networks employed in the model was $N_l \times N_u + N_{ou}$, 
comprising $N_l$ layers and $N_u$ units (neurons) per layer, followed by $N_{ou}$ output units.
Both gain sub-models utilized the architecture $N_l \times N_u + N_{ou} = 8 \times 32 + 1$, 
while the placement correction model employed $N_l \times N_u + N_{ou} = 1 \times 1 + 2$.
Tests conducted with smaller architectures demonstrated a diminished convergence efficiency. Conversely, 
tests conducted with larger architectures revealed a negligible impact on the final RMSD.
Each of the sub-models generated an output ($O_{\Delta E}, O_{E_{res}}, O_{\delta y}, O_{\delta\theta}$) within the 
range $[0-1]$, which was subsequently 
transformed by the loss function to conform to the desired parameter range.

The distinctive feature of this approach is that the three independent sub-models share a common 
loss function. This function monitors the progress of training, tracking the degree of error in the minimization 
process and enabling the derivation of gradient feedback for the models necessary for their convergence.

\subsubsection{Sub-models' inputs}
\label{sec:SubInputs}

The event-by-event inputs to the sub-models are depicted in Figure~\ref{fig:MODEL} and 
can be summarized as follows:

\begin{itemize}
\item $\Delta E$ gain: strip number $k_{\Delta E}$,
\item $E_{res}$ gain:
 
\begin{itemize}[itemindent = -2em]
\item experimental event: strip number $l_{E_{res}}$, 
\item calibration event: random number $Rnd[0,\dots, 57]$,
\end{itemize}

\item $\delta y, \delta\theta$: random number $Rnd[0,\dots, 100]$.
\end{itemize}

\subsubsection{Sub-models' output ranges}

\label{sec:SubRange}

The output range of the sub-model can be summarized as follows:

\begin{itemize}
\item $\Delta E$ gain: individual, gain obtained from $\alpha$ calibration $\pm 2~\%$,
\item $E_{res}$ gain: common, rough estimated gain $\pm 10~\%$,
\item $\delta y, \delta\theta$: $[-4,4]$~mm, $[-4,4]$ degrees,
\end{itemize}

\subsubsection{Loss's function inputs}
\label{sec:LossInputs}

The event-by-event inputs to the loss function can be summarized 
as follows:
\begin{itemize}
\item $\Delta E$ gain: $O_{\Delta E}$,
\item $E_{res}$ gain: $O_{E_{res}}$,
\item placement corrections: $O_{\delta y}$ and $O_{\delta\theta}$,
\item identifier distinguishing between experimental and calibration events: $TAG$,
\item mean position of the beam:
\begin{itemize}[itemindent = -2em]
\item experimental event: $(\overline{x}, \overline{y})_{b}$
\item calibration event: $(0, 0)$,
\end{itemize}

\item training target energy: 
\begin{itemize}[itemindent = -2em]
\item experimental event: the excitation energy\\ $E^*_{target} = 0$,
\item calibration event: the alpha energy $E_\alpha$,
\end{itemize}
\item raw energy $\Delta E_k^{raw}$,
\item raw energy:
\begin{itemize}[itemindent = -2em]
\item experimental event: $E_{{res}_l}^{raw}$,
\item calibration event: $0$.
\end{itemize}

\end{itemize}

\begin{figure*}[ht]\centering
\includegraphics[width=0.9\textwidth]{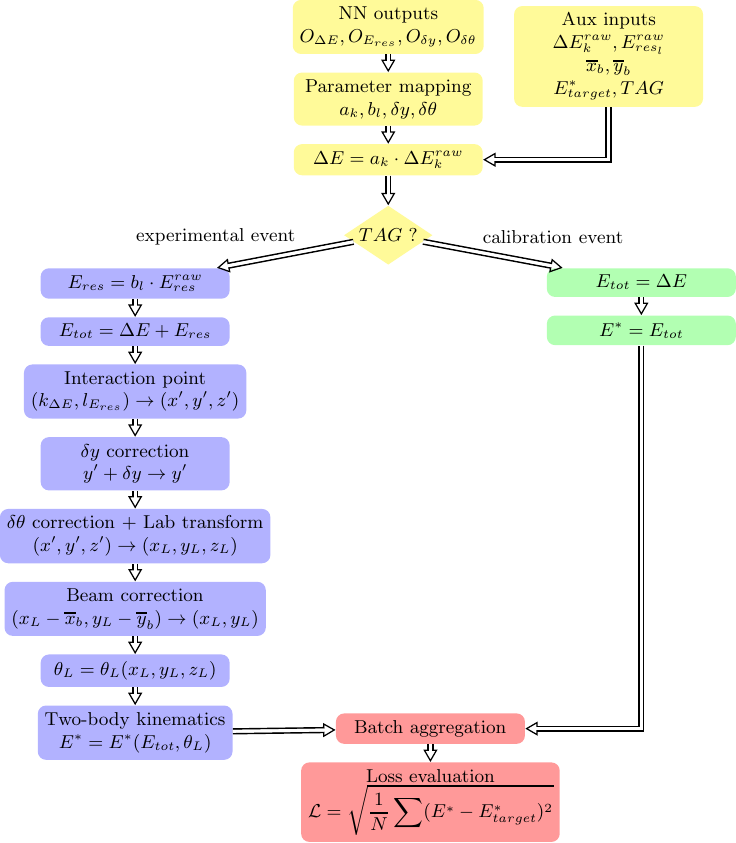}
\caption{
Algorithmic structure of the physics-informed loss function. The network outputs are mapped to physical parameters, 
followed by energy calibration. The processing then bifurcates depending on the event type: experimental events 
undergo full kinematic reconstruction, while calibration events enforce an energy anchor. Both branches contribute 
to the final RMSD loss.
}
\label{fig:loss_diagram}
\end{figure*}

\subsubsection{Algorithmic structure of the loss function}
\label{sec:Loss}

Algorithmic structure of the loss function is illustrated in Fig.~\ref{fig:loss_diagram}.
The loss function comprises three main stages: \textbf{(i)} parameter mapping, \textbf{(ii)} energy calibration, and \textbf{(iii)} event-dependent reconstruction. 
Experimental events undergo full kinematic reconstruction leading to $E^* = E^*(E_{tot}, \theta_L)$, while calibration events impose $E^* = \Delta E$.
The loss is defined as the RMSD over the batch.

\subsection{Training}
\label{sec:Train}


Since the telescopes of the PISTA array are independent, the training of the PISTA neural networks model was 
performed for each telescope individually. The datasets utilized for training the model ranged between 
$1.2-2.2\times10^6$ events. For each telescope, the pre-experiment three-alpha calibration data constituted 
one-third of each dataset.
At the beginning of the training process, the dataset was randomly partitioned into the training set ($80~\%$) 
and the validation set ($20~\%$). For each epoch, a complete iteration through the entire training dataset, 
a random rearrangement of the training dataset was performed.  The training was performed in batches of
$1\times 10^3$ events. The training convergence of the neural network 
was assessed in terms of the RMSD. The duration of a single training epoch was 
approximately $2$~s. The training and validation RMSD was evaluated every $5$ epochs ($10$~s) with early 
stopping, employing a tolerance of 0.0001 and a patience of $15$ epochs. Remarkably, the training convergence 
was typically achieved within approximately two minutes, indicating that the neural network swiftly and efficiently 
identifies the system characteristics. The model’s time efficiency can be compared to several weeks or months 
of channel-by-channel calibration.
The resulting RMSD values were ${\rm RMSD} = 0.35, 0.34$ and $0.31$~MeV 
for the telescopes located at $\phi = 90^\circ, 135^\circ$, and $180^\circ$, respectively. Training tests were conducted 
using subsampled datasets, such as employing an equal number of counts for each individual $\Delta E$ strip. 
The results exhibited stability within less than $5~\%$. Therefore, the deterministic nature of the loss function and 
the limited sensitivity to initialization ensure a high degree of reproducibility of the obtained calibration parameters.

The resulting calibration coefficients exhibited only modest random variations across the detector channels. 
For the three telescopes considered, the mean calibration coefficients were $\overline{a}=0.0134$~MeV/ch 
for the $\Delta E$ stage and $\overline{b}=0.00274$~MeV/ch for the $E_{res}$ stage. The corresponding 
channel-to-channel variations for each telescope were within $\pm 2~\%$ of $\overline{a}$ and within 
$\pm 0.5~\%$ of $\overline{b}$, respectively.
The resulting mean placement corrections were $\overline{\delta\theta}=+3.2^\circ \pm 6~\%$ and 
$\overline{\delta y}=0$~mm. These values are consistent with the detector positioning determined independently 
from the mechanical survey, providing an additional validation of the calibration procedure.

\begin{figure*}[ht]\centering
\includegraphics[width=0.45\textwidth]{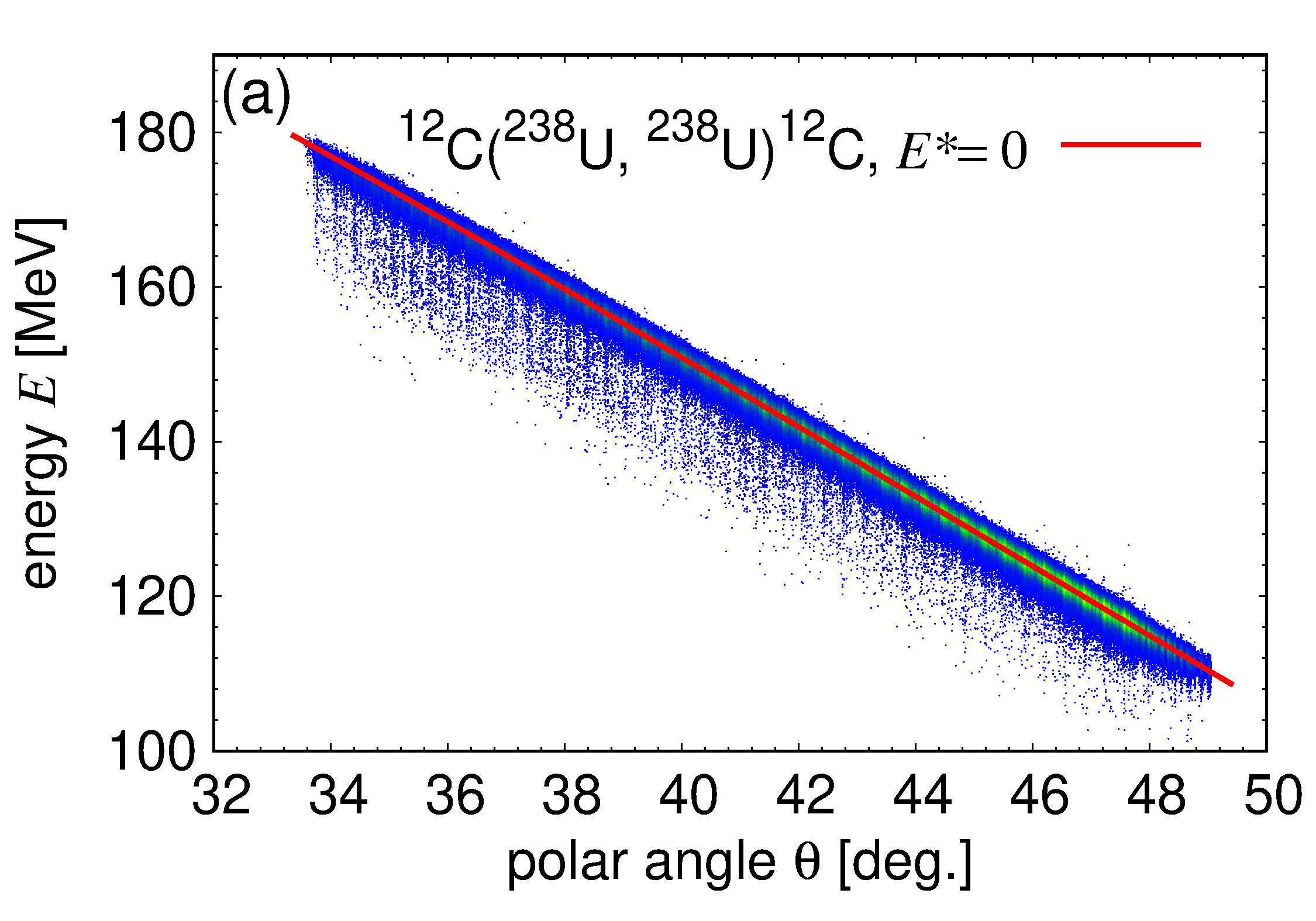}
\includegraphics[width=0.45\textwidth]{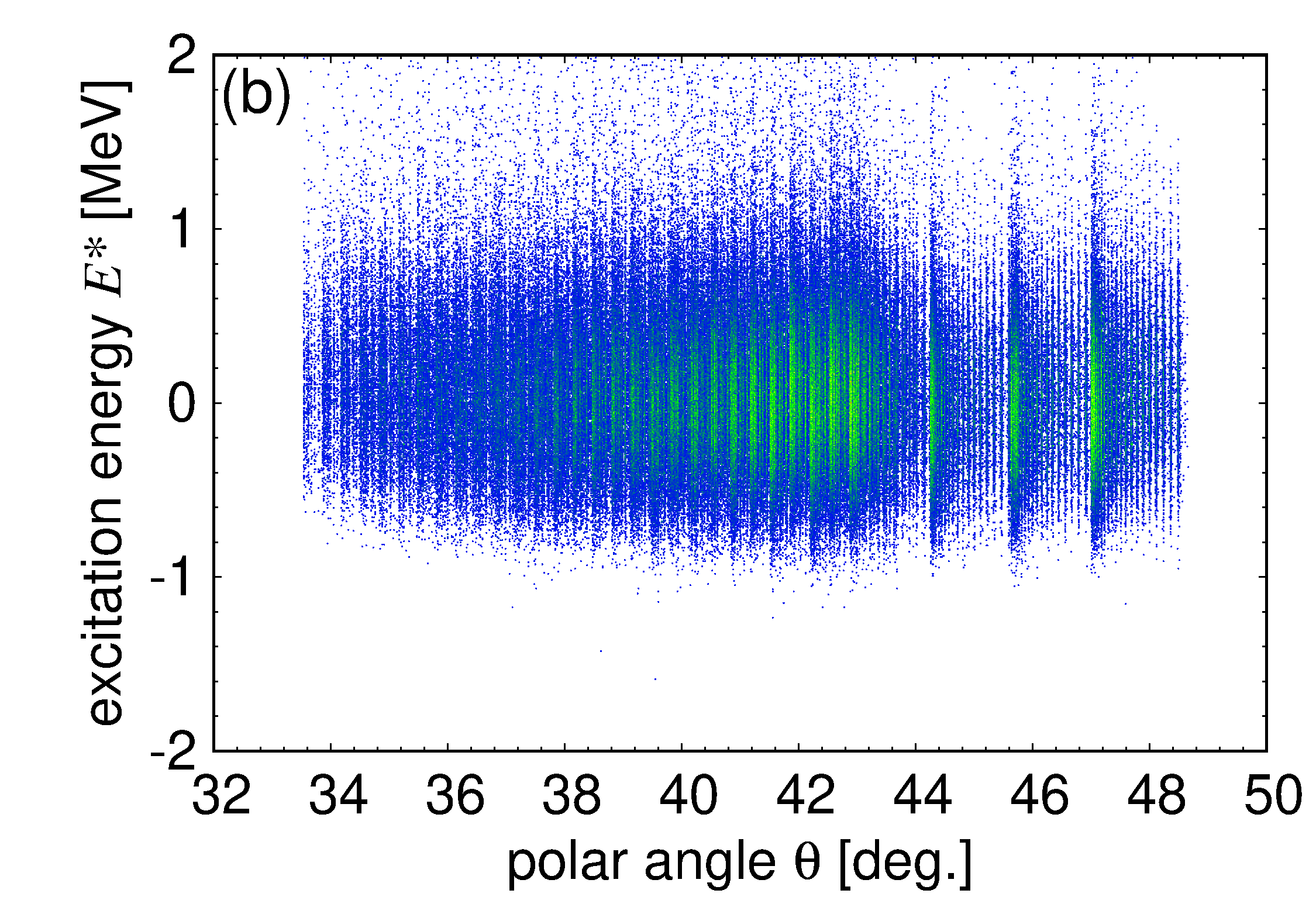}
\includegraphics[width=0.45\textwidth]{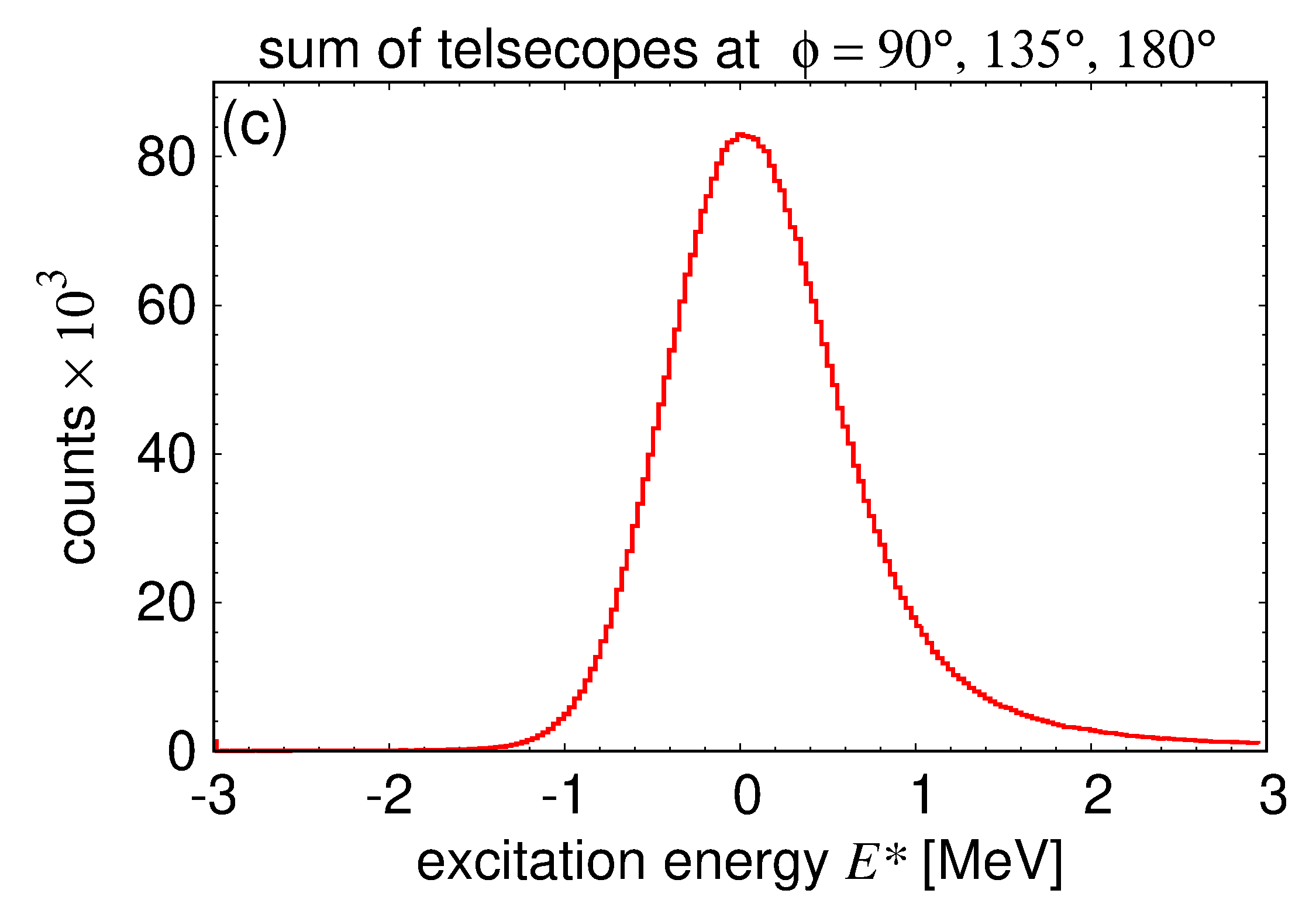}
\includegraphics[width=0.45\textwidth]{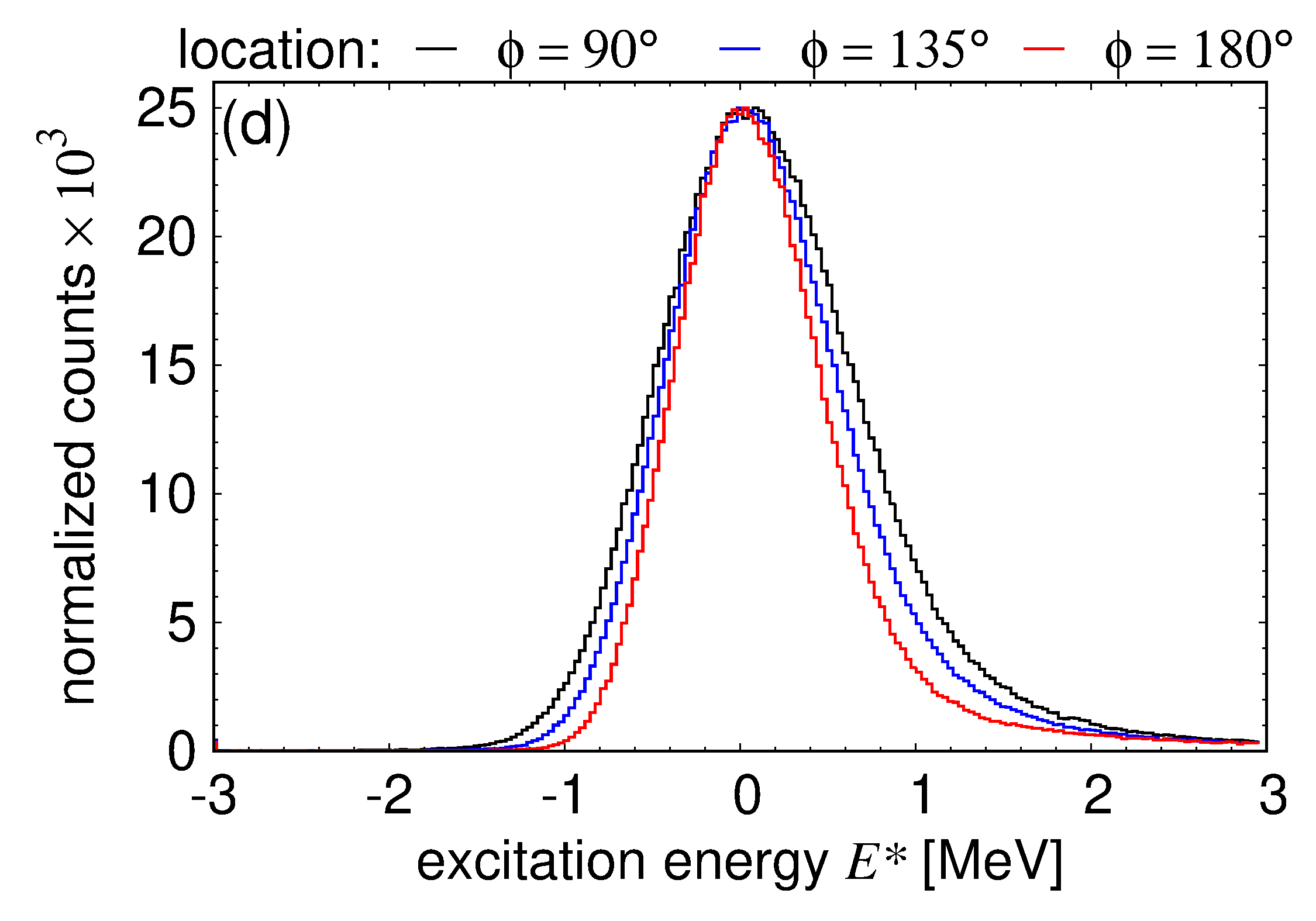}
\caption{PISTA neural network calibration results:
Two-dimensional correlation plots: 
(a) The energy ($E_{tot} = \Delta E + E_{res}$) as a function of the polar 
angle ($\theta$) for elastic scattering of $^{238}$U on $^{12}$C recorded in the telescope situated 
at $\phi = 135^\circ$. The solid red line represents the corresponding calculated correlation, leading 
to an excitation energy $E^* = 0$~MeV.
(b) The reconstructed excitation energy ($E^*$) as a function of the polar 
angle ($\theta$) for elastic scattering recorded in the telescope situated at $\phi = 180^\circ$. 
One-dimensional spectra of the reconstructed excitation energy ($E^*$):
(c) For three telescopes located at $\phi = 90^\circ, 135^\circ$, and $180^\circ$ and 
(d) For each of the three telescopes, located at $\phi = 90^\circ$ (black), $135^\circ$ (blue), and $180^\circ$ (red), 
normalized to the same peak height.
\label{fig:RES1}}
\end{figure*}

\section{Physics validation results}
\label{sec:results}

The performance of the resulting calibration is assessed through complementary physics-based observables. 
First, using the $^{12}$C data employed for training, the reconstructed elastic-scattering kinematics are used 
to verify that the correlation between the total energy and the scattering angle follows the expected two-body 
kinematics and that the reconstructed excitation energy is independent of the scattering angle. Furthermore, 
the excitation-energy resolution is verified to be smaller than the kinematic variation across one detector strip. 
In the next step, the full data set, including all nuclei observed in PISTA, is used for an independent post-calibration 
verification of the particle-identification capability through the calibrated $\Delta E$--$E_{\rm res}$ correlations.

\subsection{Excitation energy}
\label{sec:ExcEn}

\begin{figure*}[th]\centering
\includegraphics[width=0.45\textwidth]{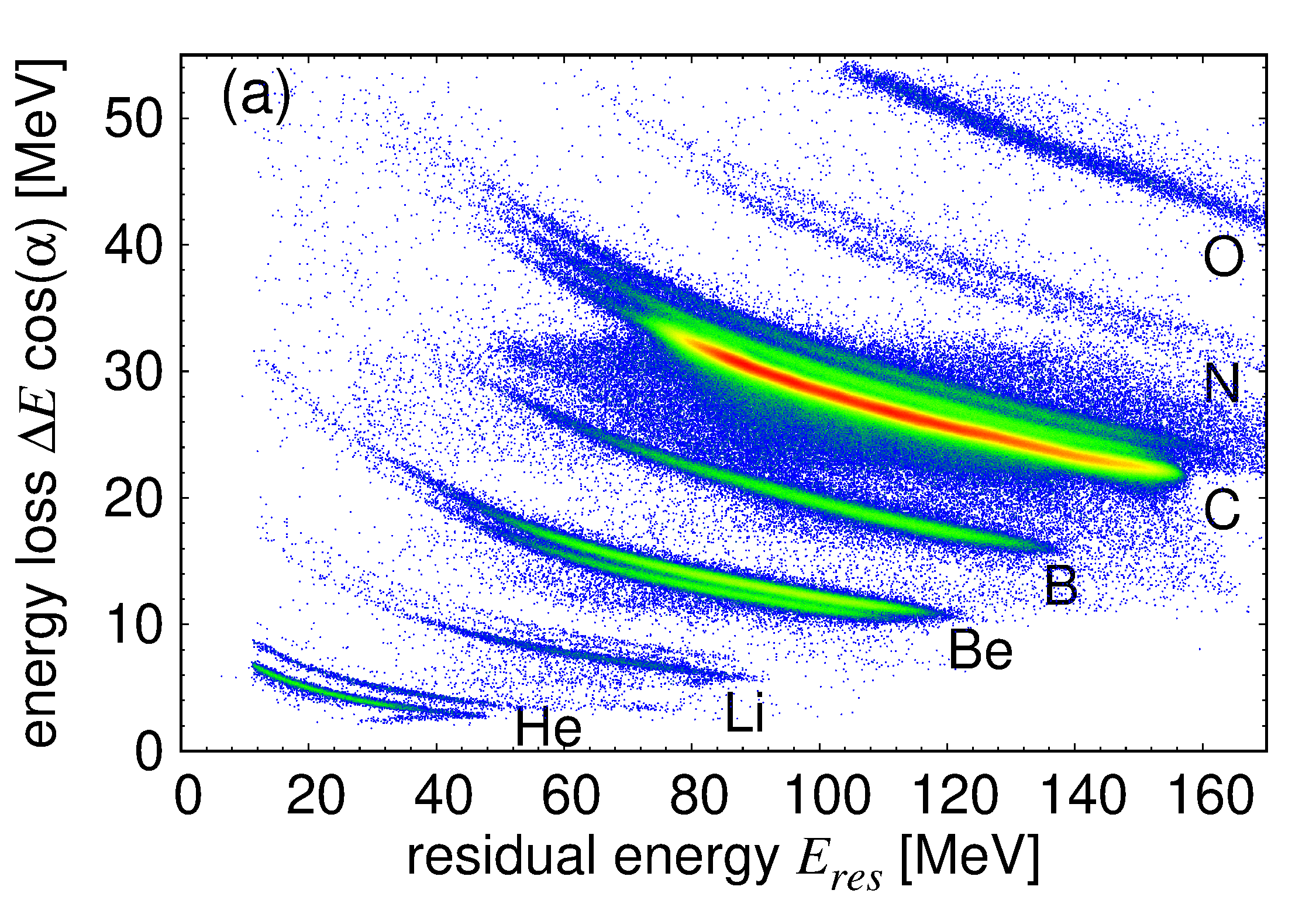}
\includegraphics[width=0.45\textwidth]{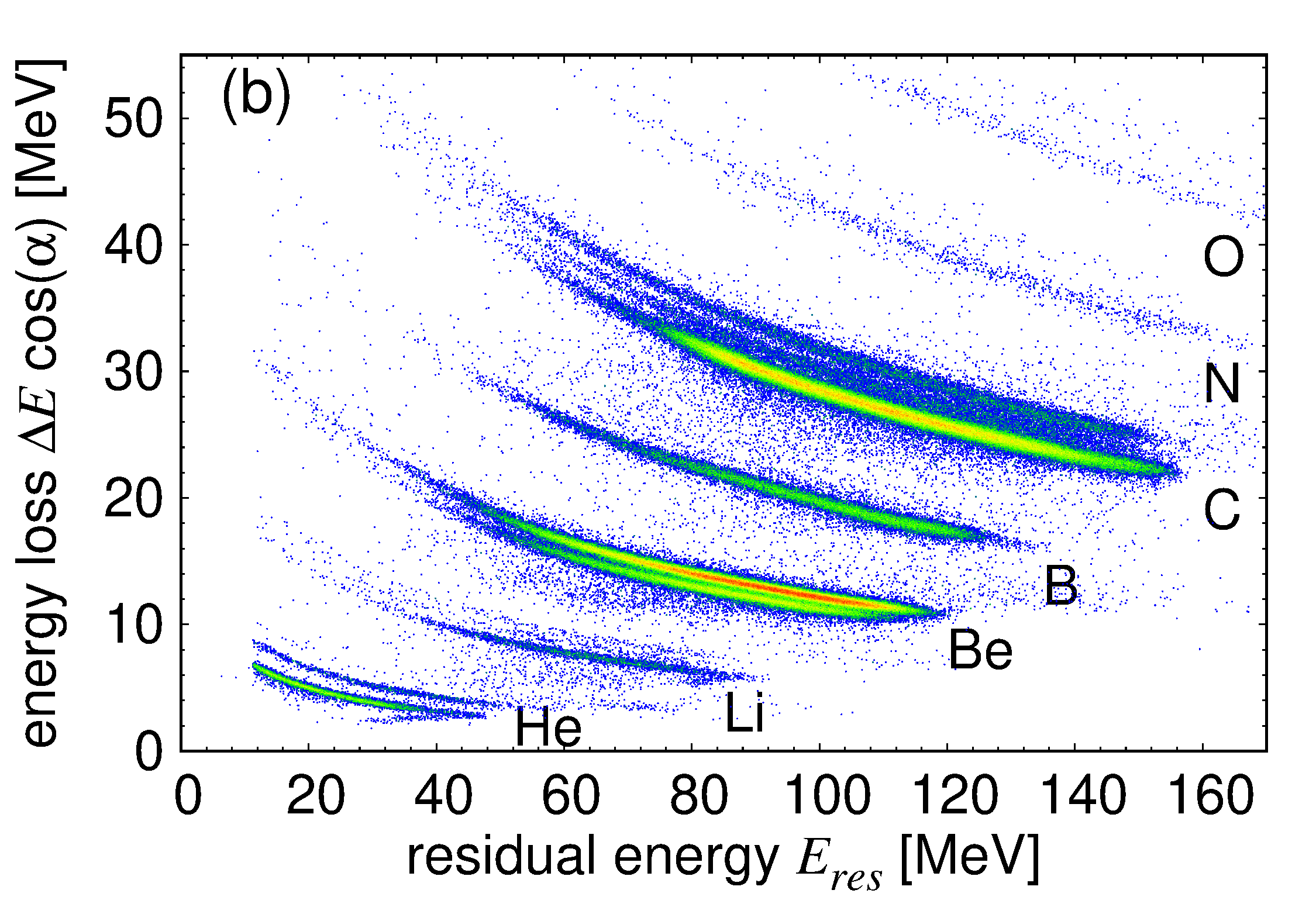}
\includegraphics[width=0.45\textwidth]{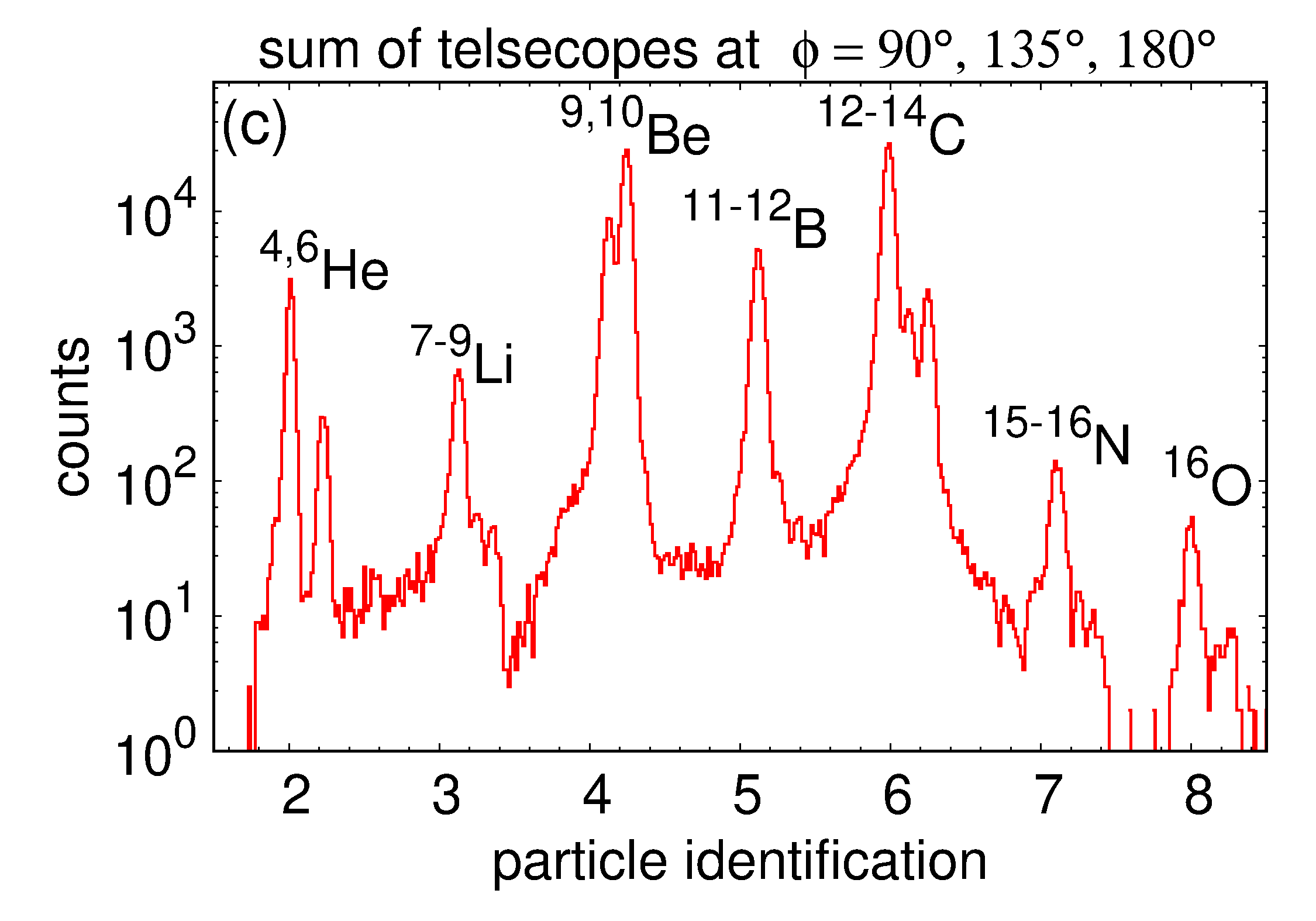}
\includegraphics[width=0.45\textwidth]{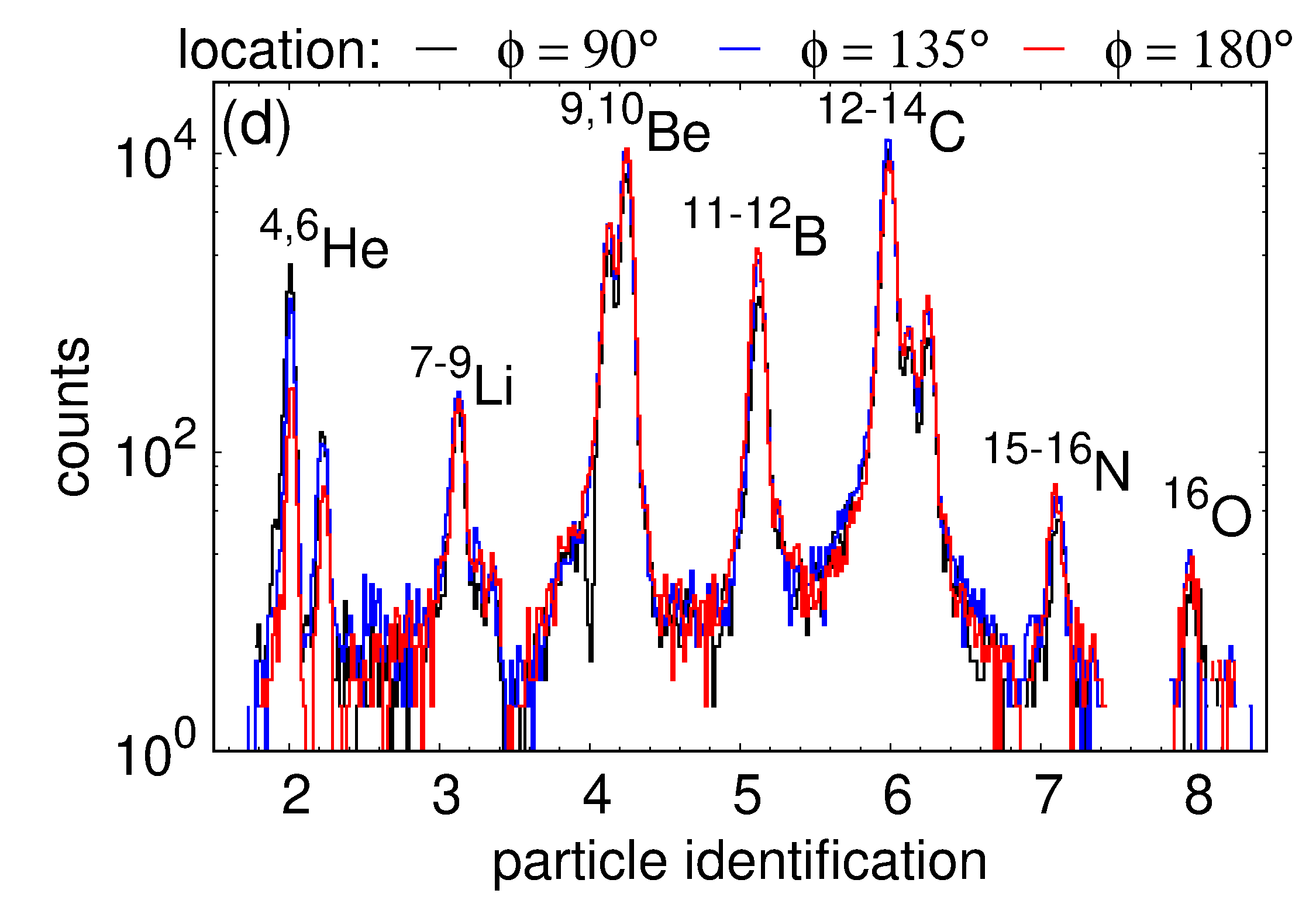}
\caption{Particle identification:
Two-dimensional energy correlation plots between the energy loss ($\Delta E$cos($\alpha$)) and the residual 
energy ($E_{res}$) for three telescopes located at $\phi = 90^\circ, 135^\circ$, and $180^\circ$:
(a) The spectra without requiring a coincidence with VAMOS++ and (b) The spectrum obtained in coincidence with any ion 
detected in VAMOS++ spectrometer. The angle ($\alpha$) is the angle between the particle's trajectory and the detector's normal.
One-dimensional spectrum of the particle identification 
(c) For three telescopes located at $\phi = 90^\circ, 135^\circ$, and $180^\circ$ and
(d) For each of the three telescopes, located at $\phi = 90^\circ$ (black), $135^\circ$ (blue), and $180^\circ$ (red).
\label{fig:RES2}}
\end{figure*}

Figure~\ref{fig:RES1}(a) depicts a two-dimensional correlation chart between the total energy ($E_{tot}$) and 
the polar angle ($\theta$) obtained for elastic scattering of $^{238}$U on $^{12}$C. The data were recorded 
by the telescope positioned at $\phi = 135^\circ$. This correlation is compared to the calculated correlation 
indicated by the red line, assuming that the total excitation of the system is $E^* = 0$~MeV. One can observe 
in the figure a perfect matching of the experimental data with the predicted curve. The pixelization observed for 
the polar angle ($\theta$) is due to the assumption that the nucleus interaction with the detector occurred at the 
center of the corresponding strips.

Figure~\ref{fig:RES1}(b) presents a two-dimensional correlation between the total excitation energy of the system 
($E^*$) and the polar angle ($\theta$) obtained using the two-body kinematics. 
The data were recorded by 
the telescope positioned at $\phi = 180^\circ$. The discontinuity in intensity, for $\theta > 43^\circ$, is due
to the absence of certain groups of strips in this telescope. 
For elastic scattering, the physical excitation energy of the target-like nucleus is $E^*=0$. The reconstructed $E^*$ 
distribution is therefore expected to be centered around zero, while its finite width reflects the combined effects 
of the detector energy and angular resolutions, energy losses, beam characteristics, and other experimental uncertainties. 
The excitation-energy width used in the calibration thus provides a measure of the quality of the reconstructed kinematics 
rather than a physical excitation-energy range of the elastic-scattering events.
The observed asymmetry in event density for the excitation 
energy $E^*$, extending towards $2$~MeV arises from the contribution of the inelastic process that leads to the 
excitation of the $^{238}$U projectile since the first excited state of $^{12}$C is located at about $4.439$~MeV 
of excitation energy. This asymmetry can be similarly observed in Fig.~\ref{fig:RES1}(a).

Figure~\ref{fig:RES1}(c) depicts a one-dimensional spectrum of the reconstructed excitation energy ($E^*$) for the 
elastic scattering of $^{238}$U on $^{12}$C. This spectrum represents the sum of the three telescopes located 
at $\phi = 90^\circ$, $135^\circ$, and $180^\circ$. The peak asymmetry, relative to the Gaussian shape, due to 
inelastic scattering events can be clearly observed in the figure. Therefore, to evaluate the obtained excitation energy 
resolution, a Gaussian fit is performed only for the left-hand side events. The resulting standard deviation is 
$\sigma_{E^*} = 0.436(10)$~MeV.

Figure~\ref{fig:RES1}(d) illustrates the corresponding individual spectra for each of the three telescopes.
The spectra were normalized to the same peak height to observe the evolution of the excitation energy 
width with the position of the telescopes. As depicted in the figure, the excitation energy width 
increases stepwise as the azimuthal angle of the telescope decreases from $\phi = 180^\circ$ to $\phi = 90^\circ$.
Since, in the case of elastic scattering, no fission fragment is detected in VAMOS++, the event-by-event interaction 
position on the target is not available. 
As discussed in Section~\ref{sec:PISTARES}, the excitation energy width is predominantly influenced by the beam profile. 
In the presented experiment, the estimated standard deviation of the beam profile exhibited asymmetry and can be expressed 
as $\sigma(b)_x = 350 \pm 50~\mu$m and $\sigma(b)_y = 550\pm50~\mu$m. The excitation energy width of the 
telescope positioned at 
$\phi = 180^\circ$ is primarily impacted by $\sigma(b)_x$, while the telescope located at $\phi = 90^\circ$ is influenced by 
$\sigma(b)_y$. The telescope situated at $\phi = 135^\circ$ is affected by an intermediate beam profile projection. The resulting 
individual standard deviations for the excitation energy are $\sigma(\phi = 90^\circ)_{E^*} = 0.490(10)$~MeV, 
$\sigma(\phi = 135^\circ)_{E^*} = 0.415(10)$~MeV, and $\sigma(\phi = 180^\circ)_{E^*} = 0.350(10)$~MeV. 
The values obtained in this work can be compared to those obtained from the conventional analysis method as in Ref.~\cite{BegueGuillou2026} for the same experimental dataset. The resulting individual standard deviations for the excitation energy are
$\sigma_{c}(\phi = 90^\circ)_{E^*} = 0.590(10)$~MeV, 
$\sigma_{c}(\phi = 135^\circ)_{E^*} = 0.530(10)$~MeV, and $\sigma_{c}(\phi = 180^\circ)_{E^*} = 0.410(10)$~MeV. 
 The mean improvement in terms of the obtained width is $18~\%$.
The obtained experimental values in this work are also lower than those obtained by the Monte Carlo simulation given in Section~\ref{sec:PISTARES} 
and require the reduction of the estimated beam profile width to $\sigma(b)_x = 250(15)~\mu$m and $\sigma(b)_y = 450(15)~\mu$m with 
which the simulation predicts the individual standard deviations for the excitation energy as follows: 
$\sigma_s(\phi = 90^\circ)_{E^*} = 0.489(10)$~MeV, $\sigma_s(\phi = 135^\circ)_{E^*} = 0.428(10)$~MeV, and 
$\sigma_s(\phi = 180^\circ)_{E^*} = 0.357(10)$~MeV.  The agreement between the experimental and simulated values of the excitation 
energy width obtained for the telescope positioned at $\phi = 135^\circ$, influenced by both vertical and horizontal beam profile widths, 
supports the accuracy of the estimated beam profile.

\subsection{Particle identification}
\label{sec:PID}

The presented calibration method utilizing PISTANN neural networks primarily correlates the total energy ($E_{tot}$) 
of the detected nuclei with the polar angle ($\theta$) of their trajectories through the two-body kinematics relation. 
Consequently, no requirements were incorporated into the PISTA neural networks model regarding the form of the 
energy loss ($\Delta E$) and residual energy $(E_{res})$. Therefore, it is of interest to verify the results of the calibration 
in terms of particle identification.

\subsubsection{Two-dimensional identification}
\label{sec:PID2D}

As described in Section~\ref{sec:PISTANN}, only the $^{12}$C events were used for the calibration of the 
PISTA neural networks model, while other reaction products were excluded. In the following we perform
an independent post-calibration validation of the resulting calibration by using the final optimized parameters 
in the standard data analysis of the complete dataset including all observed nuclei.

Figure~\ref{fig:RES2}(a) presents a two-dimensional correlation chart, without requiring a coincidence with VAMOS++, 
of the energy loss 
($\Delta E$cos($\alpha$)) as a function of the residual energy ($E_{res}$) for all three telescopes combined.
The angle $\alpha$ represents the angle between the particle’s trajectory and the detector’s normal. 
This angle is crucial for correcting the energy loss due to the varying effective thickness of the detector as a function 
of $\alpha$. In the figure, the correlated bands corresponding to different atomic numbers and atomic mass numbers 
exhibit a smooth evolution and are well-separated. 
The standard deviation for the energy loss $\sigma_{\Delta E}$ for $^{12}$C events was evaluated
in the range of residual energy $E_{res}:[90,150]$~MeV in $80$ steps of width $\Delta E_{res} = 0.75$~MeV.
The mean value of the standard deviation for the energy loss $\overline{\sigma}_{\Delta E} = 0.3$~MeV was determined.
The most prominent and overwhelming correlation corresponds to elastic scattering events. For completeness, 
Fig.~\ref{fig:RES2}(b) presents a two-dimensional chart similar to panel (a), but it is obtained by requiring a 
coincidence with any event in the VAMOS++ spectrometer. 

\subsubsection{One-dimensional identification}
\label{sec:PID1D}

For the purpose of convenience and precise quantification of the separation between distinct isotopes, 
the two-dimensional correlation depicted in Figure~\ref{fig:RES2}(b) can be transformed into a one-dimensional 
particle identification (PID) spectrum. Towards this objective, we adopted the methodology outlined in 
Ref.~\cite{Rejmund2025b} and utilized the neural network architecture $N_l \times N_u + N_{ou} = 8 \times 32 + 1$. 
The inputs to this network were \textbf{(i)} $\Delta E$cos($\alpha$), \textbf{(ii)} $E_{res}$, and \textbf{(iii)} telescope number. 
The network was trained on the data presented in Figure~\ref{fig:RES2}(b), which were obtained in coincidence 
with any event in VAMOS++.

Figure~\ref{fig:RES2}(c) presents the one-dimensional spectra of PID for the three telescopes. In PID units, the isotopes 
of a given element are grouped around the element’s atomic number. The corresponding isotopes are spaced by $0.12$~PID 
units, with the isotope having an equal number of neutrons and protons placed at the element’s atomic number. 
Figure~\ref{fig:RES2}(d) illustrates the one-dimensional spectra of PID for individual telescopes. It is evident from the figure 
that the individual spectra are highly aligned, and the achieved isotope separation quality is consistent across all telescopes. 
The mean value of the standard deviation for particle identification $\overline{\sigma}_{\rm{PID}} = 0.03$~PID units was 
determined. With $0.12$~PID units corresponding to one mass spacing
 $\sigma/A = 0.028$ for Beryllium and $\sigma/A = 0.02$ for Carbon, the achieved separation quality was observed.

\section{Summary and Conclusions}
\label{sec:Summary}

In this work, we have presented a fully automated, physics-informed calibration framework for highly segmented silicon 
telescope arrays, based on neural network optimization. The method formulates detector calibration as a global inverse 
problem, in which detector gains and geometrical corrections are determined simultaneously by minimizing the width 
of the reconstructed excitation energy under two-body kinematics constraints.

A key feature of the approach is the use of multiple neural network sub-models sharing a common loss function, 
in which the relevant physical and geometrical constraints are explicitly embedded. This design ensures a coherent 
and self-consistent calibration across all detector channels, while preserving the independence of the different 
calibration parameters. The method relies exclusively on experimental data and well-established physical principles, 
without requiring explicit modeling of the detector response.

The calibration framework was applied to experimental data obtained with the PISTA array in inverse kinematics transfer 
reactions. The results demonstrate excellent agreement with predicted kinematics, as evidenced by the precise reconstruction 
of the excitation energy and the high quality of particle identification. The achieved excitation energy resolution and isotope 
separation confirm the validity and effectiveness of the method. In addition, the training process was found to be both 
computationally efficient and robust, with stable results obtained across different dataset configurations.

From a methodological perspective, the present work illustrates how physics-informed machine learning can be used to 
address complex calibration problems in nuclear physics. By embedding domain knowledge directly into the optimization 
process, the approach avoids the limitations of purely empirical models while retaining the flexibility of data-driven methods.

The proposed framework is inherently scalable and can be readily extended to detector systems with higher granularity or 
more complex geometries. Furthermore, the general formulation of the method makes it adaptable to other experimental 
configurations, including cases involving different reaction mechanisms or detector technologies. Future developments may 
include the incorporation of more complex kinematical constraints, as well as the extension to multi-body reaction channels.

Overall, the presented approach provides a robust, reproducible, and efficient solution to the calibration of modern segmented 
detector systems, and constitutes a promising tool for next-generation nuclear physics experiments.

\bibliographystyle{apsrev4-2}
\bibliography{PISTA}

\end{document}